\documentclass{article}
\usepackage{spconf,amsmath,graphicx,hyperref}
\usepackage{amssymb}
\usepackage{textcomp}
\usepackage{xcolor}

\usepackage{booktabs}
\usepackage{multirow}
\usepackage{enumitem}
\usepackage{cite}

\title{SEAR: Spoofing Evidence-Grounded Audio Reasoning Benchmark for Audio Language Models}
\name{Rong Wan$^{1}$, Suliu Qin$^{2}$, Jiaxi Li$^{1}$, Wei Xie$^{3}$, Wenwu Wang$^{1}$, Xiaolong Han$^{1}$, Lu Yin$^{4}$, Xilu Wang$^{1}$\thanks{This work has been submitted to the IEEE for possible publication.
Copyright may be transferred without notice, after which this version may no longer be accessible.}
}
\address{$^{1}$ University of Surrey, $^{2}$ Singapore University of Technology and Design, \\ $^{3}$ Guangxi University, $^{4}$ Shenzhen University of Advanced Technology}
\begin{document}
%
\maketitle
\begin{abstract}
Audio language models (ALMs) are increasingly used for audio deepfake detection (ADD), yet existing benchmarks assess their verdicts or rationale plausibility without verifying the underlying acoustic evidence.
To address this issue, we first introduce spoofing evidence-grounded audio reasoning (SEAR), a four-task AQA benchmark to evaluate ALM-based ADD through acoustic evidence identification and quantification, deepfake detection, and forensic rationale generation.
We further propose a bona-fide-based acoustic evidence agent (BAEA), which equips a frozen ALM with controlled acoustic tools under \textsc{fixed} or \textsc{adaptive} evidence-acquisition policies. 
Experiments with six ALMs reveal a clear gap between plausible rationales and verifiable acoustic evidence reasoning, while BAEA-\textsc{Fixed} improves final verdicts and forensic rationales on both evaluation partitions. Controlled interventions further show that misleading evidence degrades both detection and grounding performance.
\end{abstract}


\vspace{-5pt}
\section{Introduction}
\label{sec:intro}
Audio language models (ALMs)\cite{ghosh2024gama} have increasingly been applied to audio deepfake detection (ADD), formulating the distinction between bona-fide and spoofed audio as an audio question-answering (AQA) task. Pioneering work such as ALLM4ADD investigates binary deepfake classification with ALMs \cite{ALLM4ADD}. More recent methods, including HIR-SDD \cite{HIR-SDD}, FT-GRPO \cite{FT-GRPO}, CoLMbo-DF \cite{CoLMbo-DF}, and HoliAntiSpoof \cite{HoliAntiSpoof}, incorporate chain-of-thought reasoning or forensic rationale generation to support an ALM's final verdict.

Despite this progress, it remains unclear whether ALM rationales and verdicts are grounded in acoustic anomalies that are verifiably present in the signal. Conventional ADD benchmarks, such as ASVspoof \cite{ASVspoof2019,ASVspoof2021,ASVspoof5} and CodecFake+ \cite{CodecFake,CodecFake+}, primarily evaluate the final detection outcomes. Recent reasoning-aware evaluations, such as TriDF \cite{TriDF} and HIR-SDD \cite{HIR-SDD}, additionally assess the divergence between models' rationale and human-like forensic trajectory. However, these benchmarks do not evaluate whether ALMs can identify and quantify fine-grained acoustic anomalies or use such evidence consistently in forensic rationale generation and deepfake detection. We then ask: \emph{Can ALMs identify and quantify signal-level acoustic anomalies and use this evidence to support forensic rationales and deepfake verdicts?}

To bridge this gap, we first introduce spoofing evidence-grounded audio reasoning (SEAR), a four-task AQA benchmark for evaluating such ALM-based ADD. SEAR is constructed through acoustic feature extraction with statistical analysis of 35 acoustic features, controlled AQA generation, and quality control. For the empirically observed difficulty of fine-grained acoustic evidence reasoning, we draw on tool-augmented ALMs for general audio reasoning \cite{tong2026autagentreinforcementlearningframework,
wijngaard2026audiotoolagentagenticframeworkaudiolanguage} and present the bona-fide-based acoustic evidence agent (BAEA) as a tool-augmented baseline for SEAR. BAEA equips a frozen ALM with a controlled signal-analysis tool and a reference distribution estimated exclusively from bona-fide training samples. Its \textsc{Fixed} policy ranks deviations across all 35 features, whereas its \textsc{Adaptive} policy lets the ALM itself select acoustic features before tool execution. The resulting structured information with evidence-identification and feature-quantification answers is provided to the ALM for deepfake detection and rationale generation. 
\begin{figure*}[t]
    \centering
    \includegraphics[width=1\textwidth]{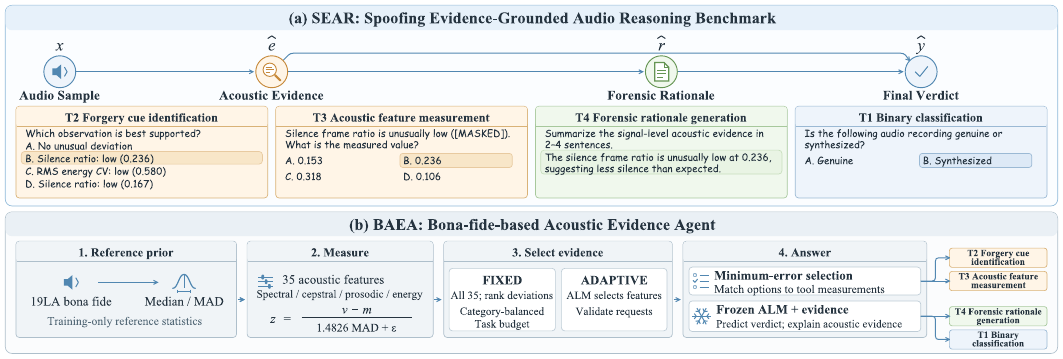}
    \caption{The proposed SEAR benchmark tasks and BAEA method.}
    \vspace{-2mm}
    \label{fig:main}
\end{figure*}

Experiments with six ALMs show a clear gap between generating plausible rationales and reasoning with verifiable evidence. On SEAR, BAEA-\textsc{Fixed} improves evidence identification and quantification, rationale grounding, and deepfake detection over its backbone. Analyses show that misleading evidence degrades both detection and grounding, further highlighting the importance of reliable acoustic evidence acquisition and use. Our contributions are threefold: 
\begin{itemize}[leftmargin=1.2em,labelsep=0.5em,itemsep=1pt,topsep=2pt,parsep=0pt]
    \item We introduce SEAR\footnote{\url{https://huggingface.co/SEAR-benchmark}}, a four-task AQA benchmark beyond binary classification for ADD, constructed via a statistically grounded acoustic-evidence pipeline.
    \item We present BAEA\footnote{\url{https://github.com/SEAR-benchmark/SEAR}}, a tool-augmented ALM that integrates controlled acoustic measurement for the tasks in SEAR without updating the backbone parameters.
    \item Experiments show that BAEA-\textsc{Fixed} improves over its backbone on SEAR, while controlled interventions reveal how evidence quality affects downstream detection and rationale grounding.
\end{itemize}
\vspace{-4pt}
\section{Our Benchmark}
\vspace{-2mm}
\subsection{Tasks Design}
\label{sec:sear}

For an audio sample $x$, SEAR evaluates ALM-based ADD from three aspects, including deepfake detection $\mathcal{D}$, acoustic evidence identification and quantification $\mathcal{E}$, and forensic rationale generation $\mathcal{R}$:
\begin{equation}
\begin{aligned}
\mathcal{D}(\hat{y},y), \qquad
\mathcal{E}(\{\hat{e}_{\mathrm{id}},\hat{e}_{\mathrm{val}}\},e^{*}), \qquad
\mathcal{R}(\hat{r},r^{*}),
\end{aligned}
\end{equation}
where $\hat{y}$, $(\hat{e}_{\mathrm{id}},\hat{e}_{\mathrm{val}})$, and $\hat{r}$ denote the ALM-generated detection, acoustic-evidence, and rationale outputs, respectively. Here, $y$ is the provided ground-truth label, while $e^{*}$ and $r^{*}$ are the reference acoustic evidence and rationale constructed by the SEAR pipeline. As shown in Fig.~\ref{fig:main}, there are four tasks designed in SEAR covering three aspects:
\begin{itemize}[leftmargin=1.2em,labelsep=0.5em,itemsep=1pt,topsep=2pt,parsep=0pt]
    \item \textbf{Deepfake verdict:} \textbf{binary classification (T1)} asks the ALM whether audio $x$ is genuine or spoofed, producing a binary verdict $\hat{y}$ evaluated against the ground-truth label $y$.
    \item \textbf{Acoustic evidence:} \textbf{forgery cue identification (T2)} asks the ALM to identify the acoustic feature that most strongly indicates spoofing, producing $\hat{e}_{\mathrm{id}}$, while \textbf{acoustic feature measurement (T3)} asks it to select the measured value of a specified anomalous feature, producing $\hat{e}_{\mathrm{val}}$. Together, T2 and T3 evaluate acoustic evidence identification and quantification against the verified acoustic evidence $e^{*}$.
    
    \item \textbf{Forensic rationale generation:} \textbf{forensic rationale generation (T4)} requires the ALM to generate an open-ended rationale $\hat{r}$ that explains its verdict using verifiable signal evidence in natural language.
\end{itemize}

\vspace{-4mm}
\subsection{Audio Sources and Acoustic Feature Extraction}

SEAR adopts two representative ADD benchmarks as its audio sources. ASVspoof 2019 LA (19LA)~\cite{ASVspoof2019} provides a controlled setting, covering six synthesis and voice-conversion attacks in its training and development sets and 13 unseen attacks in evaluation. ASVspoof 2021 LA (21LA)~\cite{ASVspoof2021} retains these 13 evaluation attacks while introducing seven codec and transmission conditions over VoIP and PSTN channels.

To obtain interpretable signal-level measurements, all audio samples are resampled to 16\,kHz and represented by 35 utterance-level acoustic features. Frame-level analysis uses an fast Fourier transform (FFT) size of 1,024 and a hop length of 256 samples. The features comprise the time-averaged first 20 mel-frequency cepstral coefficients (MFCCs) and 15 complementary measurements derived from spectral, energy, temporal, and source-related characteristics. The latter include statistics of spectral bandwidth, roll-off, centroid, and flatness; the proportion of high-frequency spectral energy; RMS energy and its temporal variation; silence and zero-crossing ratios. Together, these features characterize spectral distribution, cepstral structure, energy dynamics, temporal variation, and pitch behaviour, forming the candidate feature set for reference evidence construction.
\vspace{-9pt}
\subsection{Reference Evidence and AQA Generation}

To construct reference evidence for evaluating ALMs' answers, 35 acoustic features are adopted and screened using absolute Cohen's $d$ and folded ROC AUC, $\widetilde{A}_j=\max(A_j,1-A_j)$. 
The feature $j$ is identified as a global feature if $|d_j|\geq0.5$ and $\widetilde{A}_j\geq0.70$, where $d_j$ and $A_j$ are calculated between all spoofed and bona-fide samples. 
We identify the feature $j$ as specific to attack $a$ if its folded AUC for distinguishing spoofed samples generated by attack $a$ from the bona-fide samples satisfies $\widetilde{A}_{j,a}\geq0.80$. 

For each feature, we select the low-, high-, or two-sided anomalous region and threshold(s) that maximize balanced accuracy $\mathrm{BAcc}=(\mathrm{TPR}+\mathrm{TNR})/2$~\cite{balanced-acc}, in distinguishing spoofed from bona-fide audio.
For an audio sample $x$, feature $j$ is flagged as anomalous when its value $f_j(x)$ falls within the selected region. 
Each flagged feature yields a reference pair $(e_{\mathrm{id}}^{*},e_{\mathrm{val}}^{*})=(j,f_j(x))$. These pairs constitute $e^{*}$ and are ranked by the balanced accuracy of their corresponding rules.
If no feature is flagged, \textit{no anomaly} is assigned.

Based on the label $y$ and constructed reference evidence $e^{*}$, AQAs are instantiated. 
T1 uses $y$ as the target answer. 
For T2, the highest-ranked triggered feature, or \textit{no anomaly} if no feature is triggered, serves as the correct option and is paired with measurement-inconsistent distractors. 
T3 masks the reference measurement and constructs distractors using controlled offsets with matched precision and units. 
T4 converts the triggered findings into an evidence-grounded reference rationale. 
T1--T3 use multiple semantically equivalent templates and seeded option permutations, while T4 uses a fixed label-neutral prompt. 
Partition labels and attack identities are used only for offline reference construction and are not included in the inputs to evaluated models.
\vspace{-3mm}
\subsection{Quality Control}
Generated AQAs undergo automated checks for answer uniqueness, numerical traceability, and cross-record consistency. Six human reviewers further inspect a stratified subset of 5,400 AQAs covering labels, attack groups, feature categories, and \emph{no-anomaly} cases, assessing answer correctness, ambiguity, evidence consistency, and linguistic clarity.
\vspace{-6mm}
\section{Method}
\label{sec:method}
As shown in Section \ref{ALMs' Performance on SEAR}, many existing ALMs struggle to identify and quantify fine-grained acoustic evidence. To address this issue, we propose BAEA, by augmenting a frozen ALM with a controlled signal-analysis tool to obtain acoustic features without training, as shown in Fig.\ref{fig:main}.

\noindent\textbf{Bona-fide-based acoustic reference.} The controlled tool library uses \texttt{librosa} \cite{librosa} to compute 35 pre-registered spectral, cepstral, prosodic, and energy features. 
For feature $j$, we estimate its median $m_j$ and median absolute deviation $\mathrm{MAD}_j$ exclusively from bona fide samples in 19LA training partition. No evaluation labels, attack identities, or reference answers are used during BAEA inference. 
The robust deviation of an input audio $x$ is
$z_j(x)=(v_j(x)-m_j)/(1.4826\,\mathrm{MAD}_j+\epsilon)$,
where $v_j(x)$ is the deterministic measurement. 
The sign of $z_j(x)$ indicates the deviation direction, and $|z_j(x)|$ measures acoustic rarity; a large deviation is treated as evidence rather than direct proof of spoofing. 
Each evidence record retains the feature name, measured value, reference median, robust deviation, direction, and status. 
The ALM can invoke only whitelisted acoustic tools and cannot modify their outputs.

\begin{table*}[!t]
\centering
\caption{Zero-shot performance (\%) across SEAR tasks and partitions. Note that the training, development and evaluation sets are abbreviated as Tr, Dev and Eval, respectively.}
\label{tab:zero-shot-all-partitions}
\footnotesize
\setlength{\tabcolsep}{1.8pt}
\resizebox{0.9\textwidth}{!}{%
\begin{tabular}{l*{16}{r}}
\toprule
& \multicolumn{4}{c}{T1 (F1 $\uparrow$)} & \multicolumn{4}{c}{T2 (ACC $\uparrow$)} & \multicolumn{4}{c}{T3 (ACC $\uparrow$)} & \multicolumn{4}{c}{T4 (B-F1 $\uparrow$)} \\
\cmidrule(lr){2-5}\cmidrule(lr){6-9}\cmidrule(lr){10-13}\cmidrule(lr){14-17}
ALM & 19Tr & 19Dev & 19Eval & 21Eval & 19Tr & 19Dev & 19Eval & 21Eval & 19Tr & 19Dev & 19Eval & 21Eval & 19Tr & 19Dev & 19Eval & 21Eval \\
\midrule
Qwen2-Audio & 39.28 & 40.50 & 39.13 & 42.92 & \textbf{23.90} & \textbf{24.20} & \textbf{24.20} & 23.90 & \textbf{23.61} & 21.10 & 22.95 & 24.73 & 83.74 & 83.50 & 83.32 & 83.78 \\
Qwen2.5-Omni & 34.19 & 30.85 & 25.24 & 24.00 & 20.30 & 21.05 & 18.65 & 23.90 & 15.07 & 12.84 & 16.26 & 15.67 & 84.01 & 83.80 & 83.60 & 83.81 \\
MiniCPM-o & 13.95 & 14.00 & 13.76 & 16.06 & 19.30 & 16.65 & 14.65 & 23.70 & 22.16 & 20.23 & 22.25 & 25.77 & 84.43 & 84.19 & 83.97 & 84.16 \\
MOSS-Audio & 33.66 & 32.04 & 23.52 & 25.56 & 18.65 & 14.55 & 11.80 & 20.25 & 22.55 & \textbf{21.69} & \textbf{23.46} & \textbf{25.79} & 83.99 & 83.73 & 83.61 & 83.95 \\
\midrule
Gemini-Flash & \textbf{49.27} & \textbf{52.47} & \textbf{48.61} & \textbf{45.07} & 20.00 & 18.10 & 14.65 & \textbf{24.30} & 14.79 & 13.05 & 15.32 & 16.11 & \textbf{85.17} & \textbf{85.03} & \textbf{84.60} & \textbf{84.82} \\
GPT-Audio & 14.41 & 15.09 & 13.17 & 14.02 & 18.25 & 18.10 & 18.20 & 19.75 & 10.97 & 10.99 & 12.52 & 9.71 & 83.83 & 83.70 & 83.26 & 83.42 \\
\bottomrule
\end{tabular}}
\end{table*}

\begin{table}[t]
\centering
\vspace{-4mm}
\footnotesize
\setlength{\tabcolsep}{3.2pt}
\caption{Effects of T2/T3 context on T1 and T4.}
\label{tab:t23_transfer}
\resizebox{0.9\columnwidth}{!}{%
\begin{tabular}{@{}llccc@{}}
\toprule
Split & Model &
\shortstack{T1 $\Delta$EER\\Self/Oracle} &
\shortstack{T4 Ref-G\\Self/Oracle} &
\shortstack{T4 B-F1\\Self/Oracle} \\
\midrule
\multirow{4}{*}{19LA}
& Qwen2-Audio  & -3.85/\textbf{+1.46} & \textbf{2.542}/\textbf{3.521} & 88.23/88.14 \\
& Qwen2.5-Omni & -1.02/+0.60          & 1.915/2.550                   & 89.69/90.48 \\
& MiniCPM-o    & -1.90/-14.71         & 1.754/2.672                   & 90.19/90.19 \\
& MOSS-Audio   & \textbf{+5.77}/-1.40 & 1.851/2.929                   & \textbf{90.82}/\textbf{90.94} \\
\midrule
\multirow{4}{*}{21LA}
& Qwen2-Audio  & -11.86/-4.11         & \textbf{2.935}/\textbf{3.884} & 88.61/88.37 \\
& Qwen2.5-Omni & -1.53/\textbf{+1.62} & 2.175/2.943                   & 90.65/90.91 \\
& MiniCPM-o    & +3.23/+0.91          & 1.830/3.143                   & 91.37/91.15 \\
& MOSS-Audio   & \textbf{+3.57}/-0.56 & 1.900/3.056                   & \textbf{91.79}/\textbf{91.76} \\
\bottomrule
\vspace{-4mm}
\end{tabular}}
\end{table}

\noindent\textbf{Evidence-acquisition policies.} We instantiate BAEA with two levels of ALM control over evidence acquisition. 
One is \textsc{Fixed}, which measures all 35 features and returns the top-ranked, category-balanced deviations under a task-specific evidence budget. 
This policy provides stable evidence coverage without requiring the ALM to determine which acoustic categories should be inspected. 
The other is \textsc{Adaptive}, where the ALM selects features from spectral, cepstral, prosodic, and energy tools. 
A controller validates the request, removes invalid or repeated categories, enforces the evidence budget, and exposes only the permitted measurements. 

\noindent\textbf{Task-conditioned inference.}
The same evidence interface for all tasks is used in BAEA. For T2 and T3, it maps each option $o$ to controlled features $\mathcal{F}(o)$ and computes
\begin{equation}
d(o)=\min_{j\in\mathcal{F}(o)}
\frac{|v_j(x)-\widetilde{v}_{j,o}|}
     {|\widetilde{v}_{j,o}|+\epsilon},
\end{equation}
where $\widetilde{v}_{j,o}$ is the value stated by $o$. 
The minimum-error candidate is selected, preventing the ALM from overriding deterministic measurements.
For T1, the frozen ALM jointly uses the audio and selected evidence, with spoof scores averaged over original and A/B-swapped option orders. 
For T4, BAEA renders an immutable grounded core and retains an ALM interpretation only if it introduces no unmeasured feature, unsupported value, or attack metadata. 

\vspace{-2mm}
\section{Experiments}

\subsection{Experimental Setup}
\noindent\textbf{Models.}
We evaluate four ALMs on GH200 GPUs: Qwen2-Audio-7B-Instruct (Qwen2-Audio) \cite{Qwen2-Audio}, Qwen2.5-Omni-7B (Qwen2.5-Omni) \cite{Qwen2.5-Omni}, MiniCPM-o-4.5 (MiniCPM-o) \cite{cui2026minicpmo45realtimefullduplex}, and MOSS-Audio-8B-Instruct (MOSS-Audio) \cite{mossaudio2026}; and two proprietary models via APIs: Gemini-3.1-Flash-Lit (Gemini-Flash)~\cite{deepmind2026gemini31flashlite} and GPT-Audio-1.5 (GPT-Audio)~\cite{openai2026gptaudio15}.

\noindent\textbf{Metrics.}
T1 uses macro-F1 (F1) \cite{ALLM4ADD} and equal error rate (EER) \cite{EER}; T2--T3 use accuracy (ACC) \cite{TriDF}; and T4 uses BERTScore-F1 (B-F1) \cite{bertscore} and reference grounding (Ref-G). Ref-G is scored from 1--5 by a blinded GPT-4o-mini judge \cite{openai2025gptaudiomini}, while all other metrics are reported as percentages.

\noindent\textbf{Evaluation scale.}
2,000 audio are sampled from each partition, totaling 68,004 AQAs. The subsets retain all attack types and closely match the full class and feature distributions, with a maximum Jensen--Shannon divergence of 0.031.

\noindent\textbf{Prompt control.}
T1--T3 use three seeded, semantically equivalent templates with independently permuted options, while T4 uses a fixed label-neutral instruction. A fixed-template ablation on 200 19LA Evaluation recordings yields maximum variations of 5.7 points for T2 ACC and 3.0 points for both T3 ACC and T4 B-F1. As T1 EER varies by up to 15.0 points, we report position-ensembled EER values.

\vspace{-3mm}
\subsection{ALMs' Performance on SEAR}
\label{ALMs' Performance on SEAR}
Table~\ref{tab:zero-shot-all-partitions} reports the performance of six ALMs across the four SEAR partitions. For T1, the best F1 ranges from 45.07\% to 52.47\%, showing that direct spoofing detection remains challenging. Performance is lower on the fine-grained evidence tasks. T2 accuracy remains close to the 25\% random-choice baseline, while T3 accuracy never exceeds 25.79\%. In contrast, T4 B-F1 is consistently high, ranging from 83.26\% to 85.17\%, with little variation across models and partitions. This discrepancy shows that ALMs can generate lexically plausible rationales despite being unable to reliably identify or quantify the corresponding acoustic evidence.

We further examine whether T2/T3 answers benefit T1 and T4.
In Table~\ref{tab:t23_transfer}, Self uses model-generated T2/T3 answers, whereas Oracle uses oracle-assisted context while retaining the model answers and is not a pure gold-only upper bound. We define $\Delta\mathrm{EER}=\mathrm{EER}_{\mathrm{Direct}}
-\mathrm{EER}_{\mathrm{Context}}$, so positive values indicate improvement. 
Oracle context increases T4 Ref-G by 0.635--1.313 across all models and partitions, while B-F1 changes by at most 0.79 points. 
However, neither context
consistently improves T1, showing that accurate evidence does not automatically yield reliable detection.

\vspace{-3mm}
\subsection{Tool-Augmented Acousitc Evidence Reasoning}

Table~\ref{tab:bpae_results} compares both BAEA variants with Tool-only and vanilla Qwen2.5-Omni. Tool-only integrates its deterministic acoustic measurement for each task in SEAR without ALM inference. 
\textsc{Fixed} reduces T1 EER from 36.34\% to 24.53\% on 19LA and from 36.43\% to 33.54\% on 21LA, while improving T2/T3 and T4 Ref-G. Tool-only reproduces all T2/T3 decisions but shows dataset-dependent T1 performance, indicating that the benefit of integrating tool evidence with the ALM is domain dependent. \textsc{Adaptive} retains the same T2/T3 binding accuracy but underperforms \textsc{Fixed} on T1 and T4 in both partitions, indicating that stable feature coverage is currently more reliable than autonomous evidence selection.
\begin{table}[ht]
\centering
\vspace{-4mm}
\caption{BAEA results on SEAR.}
\label{tab:bpae_results}
\footnotesize
\setlength{\tabcolsep}{2pt}
\renewcommand{\arraystretch}{0.9}
\begin{tabular*}{0.9\columnwidth}{
    @{\extracolsep{\fill}}l l c c c c@{}}
\toprule
Split & Method
& T1 EER$\downarrow$
& T2 ACC$\uparrow$
& T3 ACC$\uparrow$
& T4 Ref-G$\uparrow$ \\
\midrule
\multirow{4}{*}{19LA}
& Vanilla
& 36.34 & 18.65 & 16.26 & 1.174 \\
& Tool-only
& \textbf{17.65} & \textbf{95.85} & \textbf{100.00} & -- \\
& \textsc{Fixed}
& 24.53 & \textbf{95.85} & \textbf{100.00} & \textbf{2.026} \\
& \textsc{Adaptive}
& 34.33 & \textbf{95.85} & \textbf{100.00} & 1.672 \\
\midrule
\multirow{4}{*}{21LA}
& Vanilla
& 36.43 & 23.90 & 15.67 & 1.196 \\
& Tool-only
& 41.64 & \textbf{84.05} & \textbf{99.99} & -- \\
& \textsc{Fixed}
& \textbf{33.54} & \textbf{84.05} & \textbf{99.99} & \textbf{1.356} \\
& \textsc{Adaptive}
& 41.58 & \textbf{84.05} & \textbf{99.99} & 1.176 \\
\bottomrule
\vspace{-5mm}
\end{tabular*}
\end{table}
\vspace{-5mm}
\subsection{Effect of Evidence Quality}
\vspace{0mm}
With BAEA-\textsc{Fixed}, we compare no, matched, and swapped evidence between recordings on the same 200 samples from each evaluation partition. 
Table~\ref{tab:evidence_intervention} reports paired improvements with 95\% bootstrap confidence intervals. 
Positive values indicate improvement over the baseline, and a star indicates that the confidence interval excludes zero.
Matched evidence improves T4 Ref-G but not T1 consistently across domains. Swapped evidence significantly increases T1 EER by 30.28 and 18.08 points and
reduces Ref-G by 0.605 and 0.755, demonstrating that misleading evidence degrades both detection and grounding.
\vspace{0mm}
\begin{table}[h!t]
\centering
\vspace{-4mm}
\footnotesize
\setlength{\tabcolsep}{1pt}
\caption{Paired evidence intervention.}
\label{tab:evidence_intervention}
\begin{tabular*}{0.8\columnwidth}{
@{\extracolsep{\fill}}llcc@{}}
\toprule
Split & Comparison & $\Delta$EER & $\Delta$Ref-G \\
\midrule
\multirow{2}{*}{19LA}
& Correct vs.\ None
& +11.11 
& +0.675$^{*}$ \\
& Swapped vs.\ Correct
& -30.28$^{*}$
& -0.605$^{*}$ \\
\midrule
\multirow{2}{*}{21LA}
& Correct vs.\ None
& -9.04
& +0.935$^{*}$ \\
& Swapped vs.\ Correct
& -18.08$^{*}$
& -0.755$^{*}$ \\
\bottomrule
\end{tabular*}
\end{table}

\vspace{-4mm}
\section{Conclusion}

We first introduced SEAR, which benchmarks ALMs across four tasks beyond binary classification in ADD. Results reveal a gap between plausible rationales and fine-grained acoustic evidence. 
We further presented BAEA, grounding the inference of an ALM by contrasting the test audio with bona fide references via controlled deterministic acoustic measurements, improving evidence grounding without model updates. 
However, BAEA relies on a relatively stable bona fide reference distribution. Future work should explore an adaptive reference modeling under evolving audio distributions.


\vfill\pagebreak

\bibliographystyle{IEEEbib}
\bibliography{strings,refs}

\end{document}